# Symmetry-Breaking Interlayer Dzyaloshinskii-Moriya Interactions in Synthetic Antiferromagnets

**Authors:** Amalio Fernández-Pacheco [1,2]*, Elena Vedmedenko[3]**, Fanny Ummelen[2,4], Rhodri Mansell[2,5], Dorothée Petit[2], Russell P. Cowburn[2].

**Affiliations:**

[1] SUPA, School of Physics and Astronomy, University of Glasgow, Glasgow G12 8QQ, United Kingdom.

[2] Cavendish Laboratory, University of Cambridge. JJ Thomson Avenue, Cambridge CB3 0HE, United Kingdom.

[3] Institute of Applied Physics, University of Hamburg. Jungiusstr 11 20355 Hamburg, Germany.

[4] Technical University of Eindhoven. 5600 MB Eindhoven, The Netherlands.

[5] Department of Applied Physics, Aalto University School of Science. P.O. Box 15100, 00076 Aalto, Finland.

\* amalio.fernandez-pacheco@glasgow.ac.uk

\*\* vedmeden@physnet.uni-hamburg.de

**Abstract:** The magnetic interfacial Dzyaloshinskii-Moriya interaction (DMI) in multi-layered thin films can lead to exotic chiral spin states, of paramount importance for future spintronic technologies. Interfacial DMI is normally manifested as an intralayer interaction, mediated via a paramagnetic heavy metal in systems lacking inversion symmetry. Here we show how, by designing synthetic antiferromagnets with canted magnetization states, it is also possible to observe interfacial interlayer-DMI at room temperature. The interlayer-DMI breaks the symmetry of the magnetic reversal process via the emergence of noncollinear spin states, which results in chiral exchange-biased hysteresis loops. This work opens up yet unexplored avenues for the development of new chiral spin textures in multi-layered thin film systems.

**Introduction**

Following its recent discovery, the interfacial Dzyaloshinskii-Moriya interaction (DMI) in ultra-thin multi-layered magnetic systems is expected to become a key constituent of future spintronic technologies (*1–3*). DMI is an antisymmetric exchange interaction emerging in systems lacking inversion symmetry that promotes chiral coupling between spins. In ferromagnetic systems, this gives rise to topological spin textures such as skyrmions and chiral domain walls, with outstanding properties to store, transport and process magnetic information (*4–9*).

Interfacial DMI in an ultra-thin ferromagnetic (FM) layer describes the coupling of spins $S_i$ and $S_j$, mediated by a paramagnetic (PM) heavy metal atom *l* in a neighboring layer (see left sketch in **Fig. 1a**), as described by the three-site Lévy-Fert model (*10*). The DMI energy per atom pair is expressed as $E_{DMI} = \boldsymbol{D}_{ij} \cdot (\boldsymbol{S}_i \times \boldsymbol{S}_j)$, where $\boldsymbol{D}_{ij}$ is the Moriya vector, whose direction is dictated by symmetry rules (*11*). This interaction favors one sense of rotation of spins in the same FM layer, *i.e.* it is a chiral intralayer interaction. Together with the vast research in FM systems, DMI plays an important role in the emerging field of antiferromagnetic (AF) spintronics (*12*). For instance, the AF interlayer coupling in between FM layers can stabilize skyrmions without the need of an external field in synthetic AFs (*13*), and interfacial DMI has been detected across a FM-AF interface (*14*), opening new opportunities for spin-torque ultra-fast switching (*15*).

Very recently, the existence of a non-negligible interlayer (IL) DMI between neighboring FM layers separated by a spacer has been predicted (*16*). Following analogous arguments as for intralayer-DMI, an IL-DMI will lead to the chiral coupling of spins of different FM layers via PM atoms located in an IL between both FMs (see right sketch in **Fig. 1a**). However, due to the rapid decrease of the DMI interaction with distance, as well as the need for the correct crystallographic symmetry, this effect has not been experimentally observed (*17*). Here, we report the experimental observation of a room-temperature chiral exchange bias in synthetic antiferromagnetic bilayers formed by one hard perpendicular layer and a soft canted layer, under in-plane magnetic fields. The thickness range where the bias is observed, as well as its symmetry, agree with the existence of a non-zero IL-DMI. The bias magnitude scales with the effective macrospin canting angle of the magnetization, evidencing that a delicate balance between magnetic energies is required for this effect to manifest. Atomistic Monte Carlo (MC) simulations that include this interaction are in good quantitative agreement with experiments, and reveal the emergence of noncollinear spin states as the mediating mechanism behind the magnetic symmetry breaking. The experimental discovery of the IL-DMI reported here opens a new route to investigate novel chiral spin textures in synthetic antiferromagnets, of great interest for future spintronic technologies.

**Synthetic antiferromagnets with canted magnetic states**

As predicted in ref. (*16*), IL-DMI should result in chiral coupling between two FM layers separated by a spacer. However, this effect, if present, is expected to be weak in comparison to other magnetic interactions, and to emerge only in FM layers with spin configurations deviating from a perfect FM alignment (this point is described in further detail later in the manuscript). To investigate this interaction, we have studied synthetic antiferromagnetic (SAF) bilayers as depicted in **Fig. 1b**. These are formed by two ultra-thin magnetic layers made of Co and CoFeB, with a heavy metal (Pt) on both sides of the two layers providing perpendicular magnetic anisotropy (PMA) and acting as a source of interfacial DMI. A Ru spacer separates both layers, coupling them antiferromagnetically via Ruderman-Kittel-Kasuya-Yosida (RKKY) interactions (*18*). The Pt layers are also used to tune the magnitude of the effective RKKY coupling (*19*). The SAF is magnetically asymmetric, with bottom Co and top CoFeB layer thicknesses respectively below and above their corresponding spin reorientation transition (SRT) thickness. This, added to the RKKY AF coupling, leads to canted magnetization states at remanence and during switching (*20*). More specifically, the samples studied here are chosen such that the bottom Co layer is significantly thinner than its SRT, *i.e.* this layer is magnetically hard, with its magnetization strongly out-of-plane (z-direction). On the contrary, the top CoFeB layer is slightly thicker than its SRT thickness, with a shape anisotropy moderately larger than its PMA (**Supplementary 1.1**). Thus, the CoFeB layer is a soft magnetic layer which, because of the competition between its low in-plane anisotropy and the AF coupling with the out-of-plane Co layer, presents canted magnetization configurations, *i.e.* it has a non-negligible magnetization component along both in-plane and z directions (*20*). Furthermore, the application of an in-plane magnetic field during growth breaks the symmetry during deposition, providing a moderate in-plane anisotropy along the field direction (*21*) (see further discussion about this point in **Supplementary 1.2**). The CoFeB in-plane easy axis is referred to as the x-direction in the manuscript. **Fig. 1b** schematically shows a typical remanent state of the system in a macrospin approximation, where $\theta$ is the effective canting angle of CoFeB, formed by the magnetization vector of this layer and the substrate normal.

**Interlayer-DMI modelling**

To estimate the IL-DMI strength, the three-site model (*10*) is applied to our system, represented as three layers arranged in an hcp stacking, with two magnetic atom layers separated by a distance $t_{IL}$ from each other by one layer of non-magnetic atoms (see **Fig. 1c**). The microscopic intralayer and interlayer DMI vectors $\boldsymbol{D}_{ij}$ are then analytically calculated (*10*), considering only next nearest neighbor FM and nearest neighbor PM atoms, as detailed in **Supplementary 2.1**. **Fig. 1c** shows the six non-zero resulting $\boldsymbol{D}_{ij}^{(Co/Pt/CoFeB)}$ vectors corresponding to the bonds connecting the central bottom Co spin $i$ and the six outer CoFeB spins $j$ of the top hexagon. The $\boldsymbol{D}_{ij}^{(Co/Pt/CoFeB)}$ vector corresponding to the interaction between both central atoms at top and bottom hexagons equals zero when computed across the three nearest neighbor impurities. From these calculations, the IL-DMI strength $|\boldsymbol{D}_{ij}^{(Co/Pt/CoFeB)}|$ is ≈ 0.02-0.03$V_1$, where $V_1$ is the so-called spin-orbit parameter of the material defining the magnitude of the $\boldsymbol{D}_{ij}$ vectors (*10, 16*). For FM/Pt interfaces, $V_1^{(FM/Pt)}$ ≈ 6.4 meV/atom (*10*), of the same order of magnitude as the direct exchange interaction of Co, $J^{(Co)}$ (*22, 23*). Hence, $|\boldsymbol{D}_{ij}^{(Co/Pt/CoFeB)}|$ ≈ 0.1-0.2 meV/atom, about one order of magnitude smaller than typical values reported for the intra-layer DMI (*24*).

We illustrate the effect of this interaction in the magnetic configuration of a bilayer SAF by depicting the ground state for the hcp unit cell in **Fig. 1d**, in the case where IL-DMI is the only (intra- or inter- layer) exchange coupling interaction considered (direct exchange coupling, intra-layer DMI and RKKY are excluded), and for large in-plane CoFeB and out-of-plane Co anisotropies. The figure shows how a strong IL-DMI with positive $\boldsymbol{D}_{ij}^{(Co/Pt/CoFeB)}$ results in an anticlockwise rotation between Co and CoFeB spins along the z-direction -from bottom to top- for spins in the same row, and clockwise for spins in adjacent rows (see **Supplementary 3** for details). This creates an alternating configuration of spins in both top and bottom layers along the x-direction (the direction of the in-plane anisotropy), as illustrated in Fig. **1e**, where an extended top view of the resulting hexagonal lattice is shown.

**Chiral exchange bias due to the Interlayer-DMI**

The presence of IL-DMI has been experimentally investigated in our SAFs by making use of a magnetic field protocol that exploits the different reversal behavior of the two layers under vector magnetic fields (*20, 25*). First, a strong unipolar -either positive or negative- (~0.4 T) $B_z$ field is applied, saturating both layers, and defining the magnetic state of the Co layer for the rest of the field sequence. This field is then set to zero, leading to a canted CoFeB layer at remanence. This initialization is followed by a moderate bipolar oscillating in-plane field (-30 mT $< B_x <$ 30 mT), applied while measuring the reversal of the CoFeB layer. This magnetic field sequence is thus a minor loop used to probe the reversal of the canted free layer, while the out-of-plane layer remains fixed along the z-direction. **Figs. 2a-d** shows experiments for one of the samples under investigation following this field sequence, where both $M_z$ (polar MOKE) and $M_x$ (longitudinal MOKE) components of the magnetization are probed as a function of $B_x$ (**Supplementary 4.1**). Importantly, the hysteresis loops associated to the CoFeB layer reversal are shifted by $B_{bias} \approx \pm 1.1$ mT for the two possible Co orientations. Analogous (bulk) vibrating sample magnetometer (VSM) measurements with two sets of perpendicular pick-up coils complement these measurements (**Supplementary 4.2**). Additional control experiments where substrate plane and magnetic field become purposely misaligned, give further evidence of the asymmetric switching of the layers, and discard any non-negligible $B_z$ offset as the source of the bias (**Supplementary 5**). Moreover, we have experimentally measured the dependence of the bias field with the angle formed by the in-plane field with the x-direction. A cosine angular dependence is observed as a function of this angle, with a maximum value when the field is applied along the x-direction, tending to zero bias for fields along the y-direction. A unidirectional bias is a consequence of the chiral character of the IL-DMI and the symmetry breaking granted by the IP anisotropy of the CoFeB layer (**Supplementary 1.2**).

To understand the experimental results, we have performed MC simulations using the atomistic model described in **Fig. 1c** (**Supplementary 2.2**). The complex polycrystalline and amorphous crystallographic structure of the sputtered layers, added to unknown spin-orbit parameters, makes it challenging to estimate the DMI values of the samples. This is added to the fact that $V_I$ will have different values for Co/Pt, Pt/CoFeB and Co/Pt/CoFeB interfaces. To incorporate realistic values in the simulations, first, we have compared sets of $M_z(B_z)$ experimental results for different thicknesses (1.7-2.2 nm) with MC simulations that incorporate analytically-calculated intra- and inter- layer DMI vectors (**Supplementary 6**). From this procedure, we estimate the spin-orbit parameters $V_I$ corresponding to the different SAF interfaces that reproduce most accurately our measurements. This also allows us to associate an effective CoFeB thickness $t$ for each sample, given by the $|V_I^{(Pt/CoFeB)}/V_I^{(Co/Pt)}|$ ratio. The estimated spin-orbit parameters are then used in subsequent MC simulations (**Figs. 2e-f**) that replicate the experimental minor loops described before. A good qualitative agreement between experiments and simulations is observed, with simulations reproducing well both the shape of the experimental loops and the bias effect. Opposite signs for $B_{bias}$ for antiparallel Co magnetization states are also obtained. Furthermore, a good quantitative agreement is also found between experiments and simulations when estimating the effective strength of the IL-DMI (**Supplementary 7**). We therefore conclude that the perpendicular bias effect described here constitutes a fingerprint of the IL-DMI.

**Effect of the interlayer-DMI across the Spin Reorientation Transition**

We have studied the dependence of the chiral $B_{bias}$ magnitude as a function of CoFeB thickness (left and bottom axes in **Fig. 3a**) for the range 1.5 - 2.4 nm. A pronounced bias is measured for samples within an interval of 0.5 nm, with a maximum value of 4 mT at 1.7 nm. This is the regime where the CoFeB magnetization becomes canted (*20*), as illustrated by the further right axis, where the function *sin2θ*, obtained from macrospin MC simulations (**Supplementary 2.3**) presents non-zero values. *sin2θ*, where *θ* is the effective macrospin canting angle of the CoFeB (see **Fig. 1b**), parametrizes the effective degree of canting of this layer (when it is neither in-plane nor out-of-plane), showing an analogous dependence with CoFeB thickness as |$B_{bias}$|. This plot therefore shows how the chiral bias can be directly mapped by *θ*. In addition, the function plotted in nearer-right and top axes is the normalized |$B_{bias}$| extracted from MC atomistic simulations as a function of the effective CoFeB thickness *t*, corresponding to an interval of |$V_1^{(Pt/CoFeB)}/V_1^{(Co/Pt)}$| between -0.2 and +1.1 (**Supplementary 6**). An excellent agreement is observed between experiments and simulations, with both functions rising sharply after the nominal SRT CoFeB thickness, peaking at 1.7 nm, and dropping to negligible values for thicknesses above 2.2 nm, when the CoFeB becomes strongly in plane.

**Fig. 3b** schematically shows the characteristic spin configurations of the system, obtained from atomistic simulations. Three thickness ranges are distinguished: *t* < 1.6 nm, (AP), 1.6 nm < *t* < 2.2 nm (CANT) and *t* > 2.2 nm (PERP). The AP and PERP are standard spin configurations, whereas the spin state for the CANT regime is explained in next section. The figure shows how no measurable bias is observed for the AP and PERP regimes, where the CoFeB layer has a net strong PMA and in-plane anisotropy, respectively. The AP configuration obviously leads to zero IL-DMI due to both layers forming 180°, resulting in $S_i \times S_j = 0$ for all pairs. However, the net IL-DMI is also zero in the PERP configuration, despite Co and CoFeB spins forming 90°. To understand this, we recall that the IL-DMI energy $E_{DMI}^{(Co/Pt/CoFeB)}$, results from a double summation of $D_{ij} \cdot (S_i \times S_j)$ over all *i, j* atomic pairs (**Supplementary 2.1**). At the PERP regime, the cross product $S_i \times S_j$ is the same for all atomic pairs, leading to $E_{DMI}^{(Co/Pt/CoFeB)} = 0$, because the total contribution of all $D_{ij}$ vectors cancel out, following symmetry arguments (**Supplementary 2.1** and **Supplementary 3**). A measurable $B_{bias}$ is therefore only present for the CANT regime, where a small effective CoFeB anisotropy is expected to promote the emergence of effects ruled by small energy contributions, such as the IL-DMI.

**Emergence of Spin Modulations**

To understand the type of spin states present in the CANT regime and its role in the chiral bias, **Fig. 4** includes results from MC simulations for a SAF within this thickness regime. **Fig. 4a** shows snapshots during the reversal process of the CoFeB layer at different $B_x$ values, for Co pointing along the +z direction (**Figs. 2(g,h)** are the corresponding hysteresis loops). Overall, the magnetization process follows the same mechanism previously reported for this range of anisotropies and RKKY coupling (*20*), result of the competing energies present in the system: The soft layer (CoFeB) reverses back and forth under $B_x$, while the hard (Co) layer remains unchanged because of its high PMA. The AF RKKY promotes an antiparallel orientation of CoFeB and Co, leading to a peak in $M_z$ during CoFeB reversal (**Fig. 2g**). The AF RKKY also results in an incomplete in-plane saturation of CoFeB at the maximum $B_x$ applied (**Fig. 2h**). In addition, the intralayer-DMI promotes a chiral clockwise spin rotation -from left to right- across the CoFeB layer. To satisfy this requirement, the magnetization reverses via the propagation of domain walls with clockwise chirality. To achieve the same wall chirality for both branches of the hysteresis loop and keep an antiparallel alignment with Co, a domain wall is nucleated at opposite edges of the simulated area for either branch. However, none of these contributions is able to create a biased switching in extended structures and under $B_x$ only (*26*), requiring the IL-DMI as a symmetry-breaking mechanism. The reversal of CoFeB will be in reality strongly influenced by defects and inhomogeneities of the layers (*27*), and driven by domains of very small sizes for thicknesses around the SRT (*28*), making their direct observation using magneto-optical methods as those used here very challenging (*25*). Despite these, the macroscopic bias observed experimentally indicates that a clear reversal asymmetry for both branches is present.

Complementing these results, **Figs. 2i-j** show the evolution of $E_{DMI}^{(Co/Pt/CoFeB)}$ during CoFeB reversal, for the two possible z-directions of Co. Whereas standard magnetic energy terms are symmetric under inversion of $B_x$, this is not the case for $E_{DMI}^{(Co/Pt/CoFeB)}$, which presents two plateaus at moderate $B_x$ values and a biased switching. An asterisk in those graphs marks the state of the system that is energetically more favorable from an IL-DMI point of view, which is depicted in the insets of **Figs. 2f-h**. These sketches show the spin configuration for top CoFeB and bottom Co layers, where green (red) interconnecting lines indicate the pair bonds where the IL-DMI is energetically favorable (unfavorable) for that spin configuration (compare with **Fig. 1d**). The figure also depicts how unfavorable bonds cause canting of the CoFeB spins (red arrows) which become more antiparallel to Co because of the strong AF RKKY interaction.

MC simulations evidence the emergence of this type of noncollinear magnetization states, as a result of the competition between IL-DMI and RKKY coupling (**Fig. 4b**). Magnetization amplitude changes of up to 15% for $S_x$, with a period corresponding to a few atomic lattice constants, are observed in simulations, with this behavior dependent on the $|V_1^{(Pt/CoFeB)}|/|V_1^{(Co/Pt)}|$ ratio (not shown here). The relevance of noncollinear magnetic phases for symmetry breaking has already been pointed out (*33*). Here, simulations show how the magnetization modulation present during the CoFeB reversal is different for the two branches of the hysteresis loop, as a consequence of the different configuration of -energetically satisfied and unsatisfied- IL-DMI bonds for both branches (**Fig. 4(c,d)**). The emergence of noncollinear spin states subject to this asymmetric bond profile is thus the subtle symmetry breaking mechanism that leads to a chiral exchange bias for magnetization and IL-DMI energy hysteresis loops (see **Supplementary 3** for further discussion). This magnetization modulation asymmetry also manifests as other small asymmetric features in these loops. For instance, the $M_z$ peak reaches larger values for one of the two branches (**Figs. 2e, g**); this reveals spin modulations of larger amplitude, due to the IL-DMI competing less efficiently with the RKKY interaction. Whereas the IL-DMI would appear to be too weak to significantly change the intralayer magnetic ordering, due to the competition with a strong direct exchange and intralayer-DMI contributions, it can however be effective in competition with RKKY coupling, co-defining the IL ordering.

**Conclusions and Perspectives**

The bias effect reported here is robust: we have observed it using several magnetometry techniques that probe different length scales, and for other sample series similar to the ones shown in the main manuscript (**Supplementary 8**). Other perpendicular indirect exchange interactions such as the biquadratic interlayer coupling (*29*) cannot account for the chiral nature of the observed effect. A crystalline texture substantially tilted with respect to the substrate plane, a possible source of perpendicular bias (*30*), is not a relevant factor in our samples (**Supplementary 9**). Furthermore, intra-layer DMI effects leading to asymmetric magnetic hysteresis processes have been only observed in laterally-patterned nanomagnets, and require the simultaneous application of orthogonal magnetic fields (*31, 32*), in contast to our experiments.

In conclusion, we report a room temperature chiral exchange bias in ultra-thin asymmetric synthetic antiferromagnetic bilayers caused by the presence of DMI across the interlayer. The bias is observed during the in-plane reversal of a canted magnetic layer, subject to strong AF coupling to a fixed perpendicular layer. The emergence of noncollinear spin modulation, subject to different IL-DMI profiles during magnetic reversal, is behind this symmetry breaking. We expect IL-DMI to manifest as well in other ultra-thin SAF systems away from the SRT, where the magnetic reversal becomes dominated by areas with a low effective anisotropy, such as defects and layer inhomogeneities. The realization of systems integrating interlayer magnetic chiral interactions paves the way for the creation and manipulation of novel spin textures in synthetic antiferromagnetic spintronic systems (*12, 15, 34–36*).

**Acknowledgments:**

We acknowledge fruitful discussions with Nicolas Jaouen, Stefan Stanescu and Aurelio Hierro-Rodríguez, as well as experimental support from Dédalo Sanz Hernández, Alexander Welbourne, Peter Seem and Ian Farrer. **Funding:** AFP acknowledges funding from an EPSRC Early Career Fellowship EP/M008517/1, and from the Winton Program for the Physics of Sustainability. EV from Horizon 2020 research and innovation program under Grant Agreement No. 665095 (MAGicSky). DP and RPC from the Templeton World Charity Foundation. FU thanks the Erasmus Mobility program. **Author contributions:** AFP designed and carried out the experiments, grew some of the samples, analyzed the data, did the MC macrospin simulations and wrote the manuscript. EV performed the analytical calculations, did the atomistic MC simulations and analyzed the data derived from them, and wrote the manuscript. FU grew some of the samples and analyzed data. RM contributed to the experimental characterization of the samples. All authors discussed and contributed to the interpretation of the results, as well as to the writing of the manuscript. **Competing interests:** Authors declare no competing interests. **Data and materials availability:** To comply with EPSRC policy, all data associated to this publication will be available via the University of Glasgow public repository.

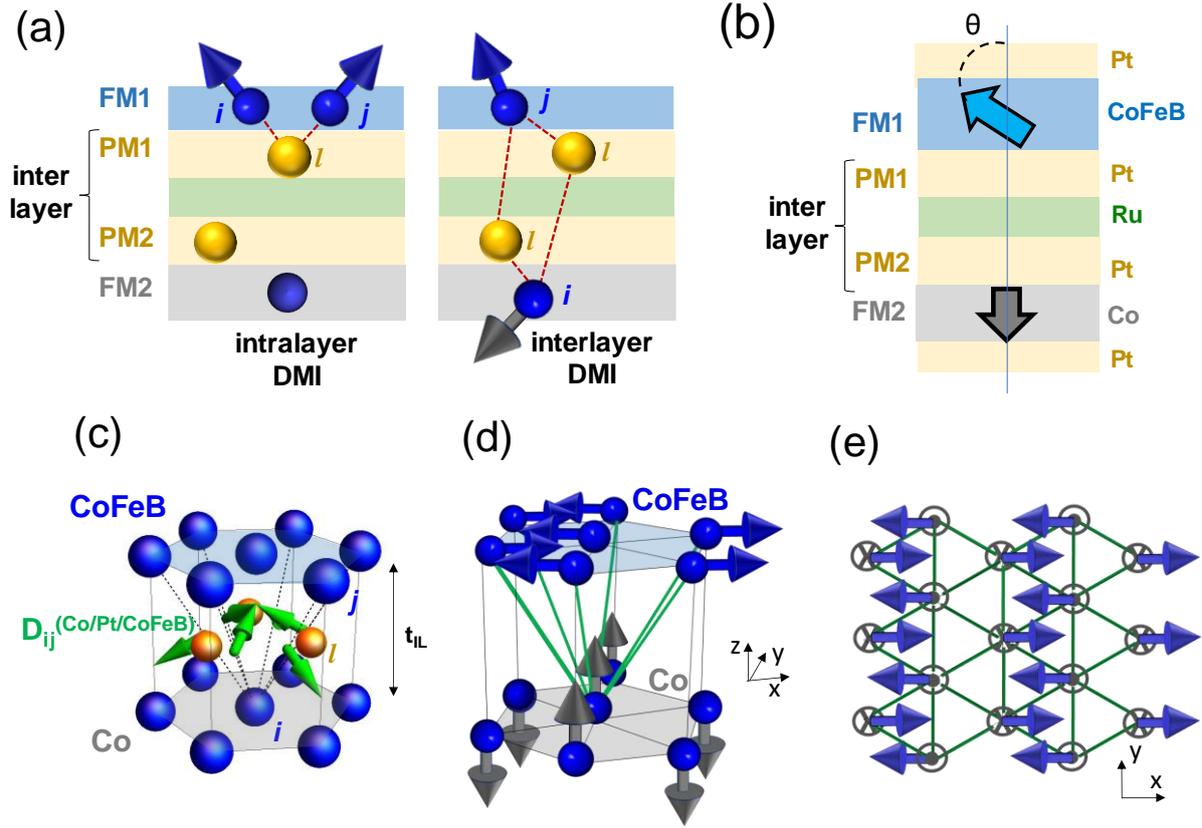

**Fig. 1. Interlayer-DMI investigations in canted SAFs.** (**a**) The presence of interfacial intralayer DMI (left) in ultra-thin FM layers results in interacting spins of the same FM layer via a PM, leading to chiral coupling and noncollinear spin configurations within the layer; the figure depicts this type of coupling for the top FM layer only. Analogously, an interlayer DMI effect (right) results in a chiral coupling between spins of two neighboring layers separated by a spacer, mediated by PM atoms. (**b**) Schematic of the types of SAFs studied and their magnetic state at remanence in a macrospin approximation: two ultra-thin CoFeB (top) and Co (bottom) layers with Pt at the interfaces, separated by Ru to create AF coupling between both FM layers via RKKY interaction. The two FM layers have different proximities to their corresponding SRT: whereas the Co layer remains out-of-plane due to its strong PMA, the CoFeB layer is just thicker than its SRT. Due to this and the AF coupling, this layer becomes canted with respect to the substrate plane. $\theta$ is the (polar) effective macrospin canting angle of this layer. (**c**) $D_{ij}^{(Co/Pt/CoFeB)}$ IL-DMI vectors (green) calculated via the 3-sites model for a Co($i$)/Pt($l$)/CoFeB($j$) trilayer with hcp structure. The distance between magnetic atoms is the interlayer thickness ($t_{IL}$). The $j$ letter in the figure denotes one of the seven next-nearest neighbours of the $i$ central bottom spin, with $l$ the corresponding PM atom for this bond included in the calculations. (**d**) Ground state spin configuration based solely on the IL-DMI, for a hcp trilayer with in-plane top and out-of-plane bottom magnetizations (no FM direct or AF RKKY exchange is considered). All green bonds connecting the middle Co to the outer CoFeB spins are IL-DMI energetically favorable. (**e**) Extended top view of the hexagonal lattice for the same ground state as in (d).

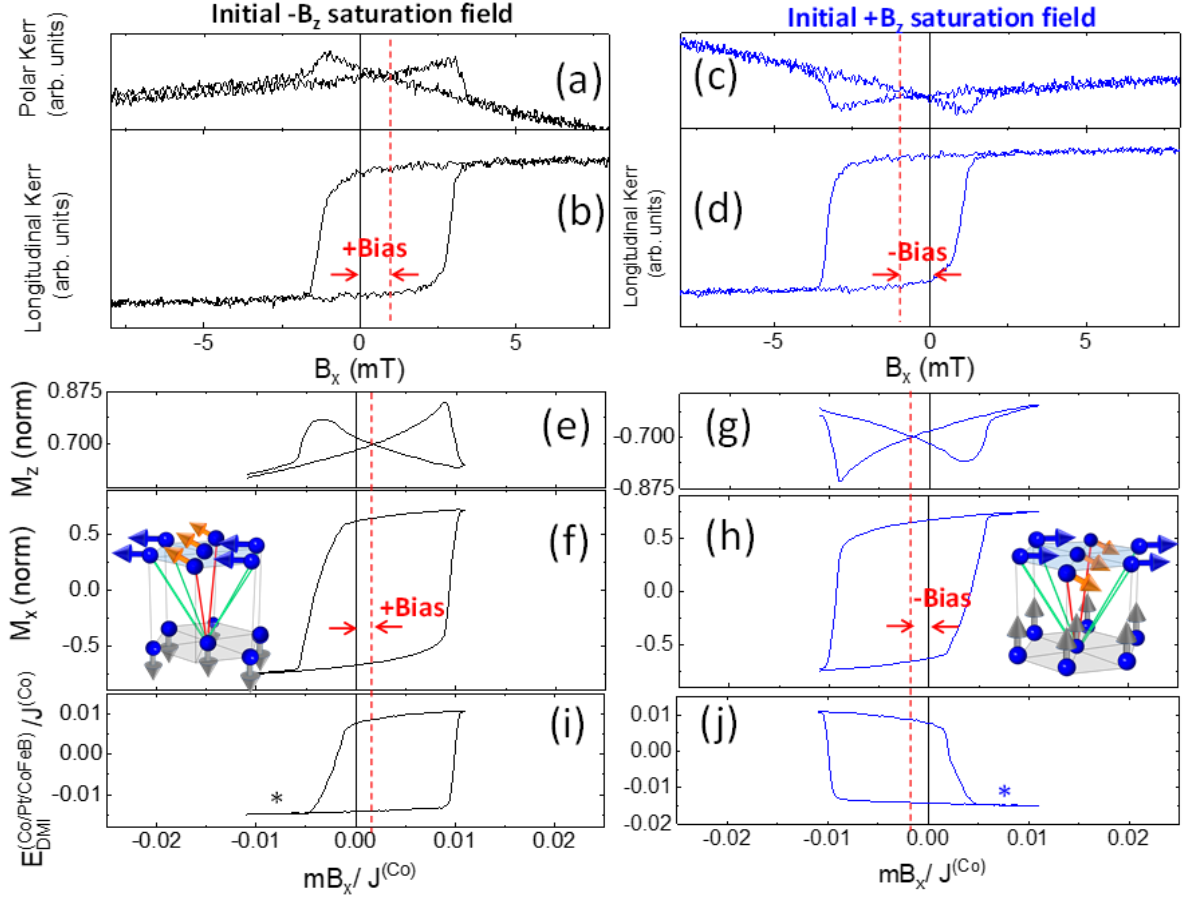

**Fig. 2. Chiral exchange bias due to the IL-DMI. (a-d)** Chiral exchange bias observed during the reversal of the canted CoFeB layer, for a sample with CoFeB thickness = 2.1 nm. The magnetization components $M_z$ (a,c) and $M_x$ (b,d) are measured by Kerr effect under $B_x$ magnetic fields, after negative (a,b) and positive (c,d) initial saturating orthogonal $B_z$ fields. The bias effect, obtained from the switching field ($M_x$) and peaks ($M_z$), is marked by a red dashed line. **(e-h)** Monte Carlo atomistic simulations reproducing the experiments, with $V_1^{(Pt/CoFeB)}/V_1^{(Co/Pt)}$ = 0.2, corresponding to a CoFeB thickness $t$ = 2.1 nm. **(i, j)** Evolution of the IL-DMI energy during the hysteresis loops; an asterisk marks the most favorable state of the two under moderate $B_x$ fields, based on this interaction. This state is sketched in the inset of (f) and (h), where red/green lines denote IL-DMI energetically unfavorable/favorable bonds connecting $j$ top outer spins to the central $i$ bottom spin. Canted spins promoted by the RKKY interaction and an unfavorable IL-DMI are colored in red. Both $mB_x$ and IL-DMI energies are normalized with respect to $J^{(Co)}$, the direct intralayer exchange energy, with $m$ the magnetic moment of the system.

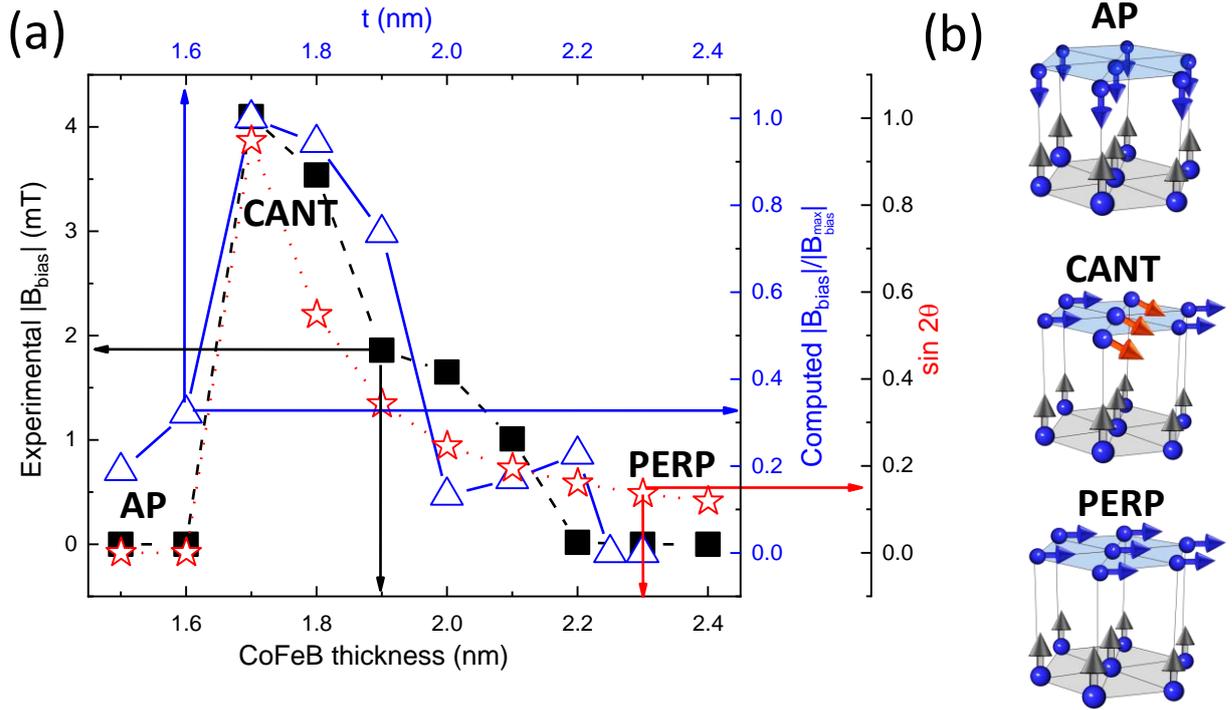

**Fig. 3. Bias field dependence with CoFeB thickness. (a) Left, bottom axes** (black squares and dashed line) are experiments, showing a peak around the SRT. **Nearer-right, top axes** (blue triangles and dash-dot line) show computed normalized bias from atomistic MC simulations, with $t$ the effective CoFeB thickness. The same behavior is evidenced for experiments and simulations. **Further-right, bottom axes** (red stars and dotted line) plots the degree of canting of the CoFeB layer as a function of its thickness, parametrized as $sin2\theta$, as extracted from macrospin MC simulations; only anisotropies and RKKY coupling interactions are considered. The magnitude of the bias is well correlated with the magnetization effective degree of canting of the CoFeB layer, revealing that a low competing effective anisotropy is necessary to observe a bias effect. **(b)** Schematics of the three types of spin configurations (AP, CANT, PERP) across the SRT. A non-zero net IL-DMI is only present for the CANT regime.

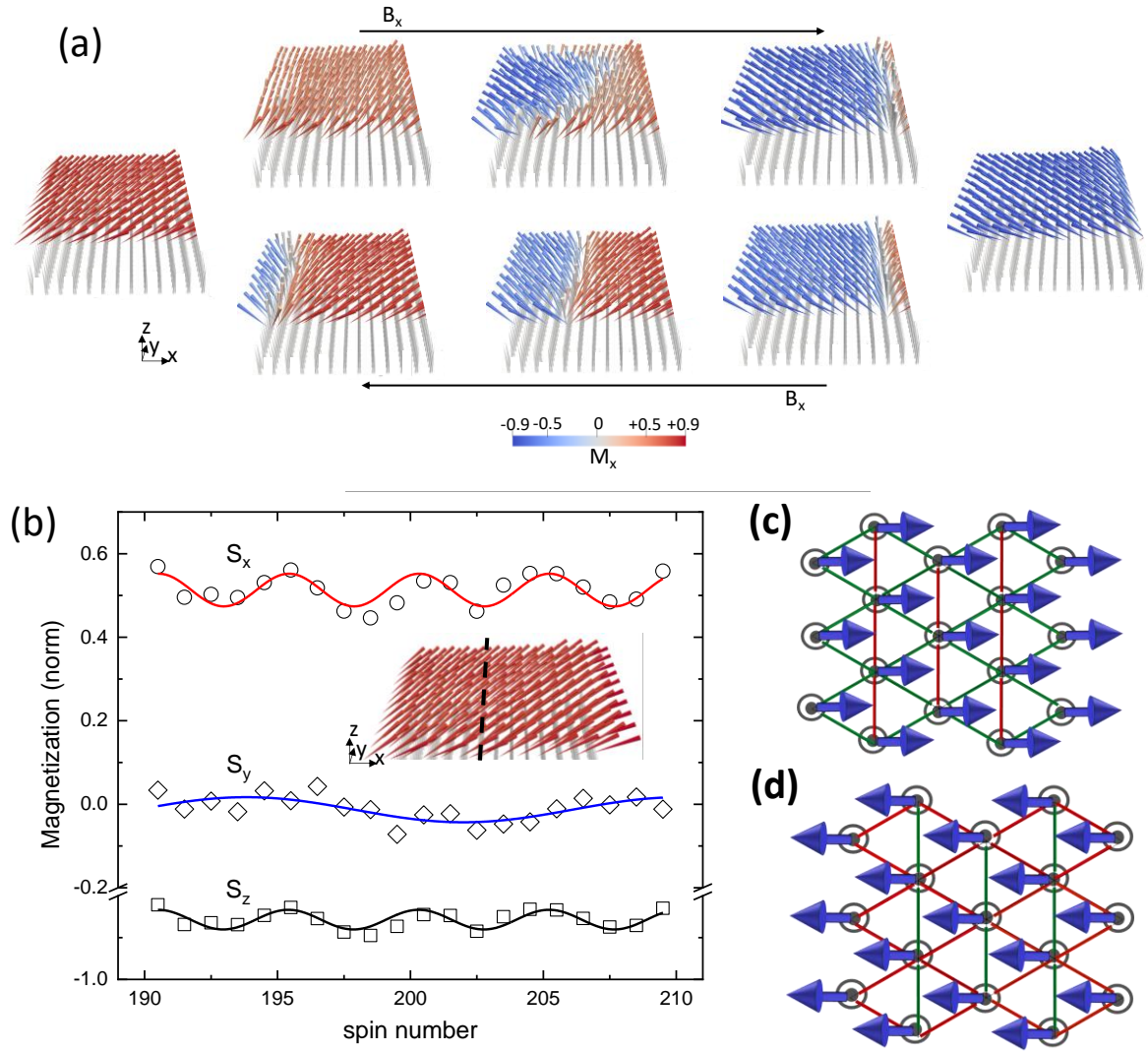

**Fig. 4. Emergence of spin modulations. (a)** Snapshots of Monte Carlo simulations at remanence for a SAF with an effective CoFeB thickness = 2.1 nm and Co pointing upwards. Forward (top) and backward (bottom) branches of the $B_x$ hysteresis loop are included. Top spins in red and blue indicate the value of $M_x$ for the top CoFeB layer during reversal. The grey bottom spins represent the Co layer along +z. The reversal process is asymmetric for both loop branches and occurs at different magnetic fields, resulting in a biased hysteresis loop. **(b)** Three components of the magnetization extracted from the simulations across the dashed line in the inset, for $B_x = 0$ and starting from negative fields. Periodic changes in the amplitude of the three components reveal the presence of spin modulations in the CoFeB layer. Different periods for the three components are observed due to their anharmonic character. **(c,d)** Top extended view of the hexagonal lattice, with bottom Co spins colored in grey and CoFeB top spins in blue. $E_{DMI}^{(Co/Pt/CoFeB)} = 0$ for both spin configurations. However, a different number and symmetry of favorable (green) and unfavorable (red) IL-DMI bonds is obtained for (b) and (c), breaking the symmetry of the system. This leads to a chiral bias when spin modulations become present during the switching of the CoFeB layer.